 \documentclass[final,5p,times,twocolumn]{elsarticle}

\usepackage{amssymb}

\usepackage{lineno}

\journal{Nuclear Physics B}

\begin{document}
\begin{frontmatter}



\title{Numerical Studies of Electron Acceleration Behind Self-Modulating Proton Beam in~Plasma with~a~Density Gradient}


\author[1]{A. Petrenko}
\author[2,3]{K. Lotov}
\author[2,3]{A. Sosedkin}
\address[1]{CERN, CH-1211 Geneva 23, Switzerland}
\address[2]{Budker Institute of Nuclear Physics SB RAS, 630090 Novosibirsk, Russia}
\address[3]{Novosibirsk State University, 630090 Novosibirsk, Russia}

\begin{abstract}
Presently available high-energy proton beams in circular accelerators carry enough momentum to accelerate
high-intensity electron and positron beams to the TeV energy scale over several hundred meters of the plasma with a density
of about $10^{15}~\mathrm{cm}^{-3}$. However, the plasma wavelength at this density is 100-1000 times shorter than the typical longitudinal
size of the high-energy proton beam. Therefore the self-modulation instability (SMI) of a long ($\sim$10~cm) proton beam in
the plasma should be used to create the train of micro-bunches which would then drive the plasma wake resonantly.
Changing the plasma density profile offers a simple way to control the development of the SMI and the acceleration of
particles during this process. We present simulations of the possible use of a plasma density gradient as a way to control
the acceleration of the electron beam during the development of the SMI of a 400~GeV proton beam in a 10~m long
plasma. This work is done in the context of the AWAKE project --- the proof-of-principle experiment on proton driven
plasma wakefield acceleration at CERN.
\end{abstract}


\end{frontmatter}


\section{Introduction}

Proton beams in the modern high-energy accelerators carry a large amount of energy. For example, the design parameters of the Large Hadron Collider correspond to the total proton beam energy of 360 MJ stored in the 2800 bunches at the top energy of 7~TeV. This is comparable to the kinetic energy of a typical fully loaded airliner (80~t) at the take-off speed of 300~km/h (280~MJ). Even the single bunch ($3 \cdot 10^{11}$ protons at 400~GeV) in the 40~year~old Super Proton Synchrotron (SPS) at CERN carries an order of magnitude more energy than the single bunch in the proposed International Linear Collider ($2 \cdot 10^{10}$ electrons or positrons at 250~GeV). This makes it possible to accelerate a substantial amount of particles to a TeV-scale energy in a single accelerating stage using a beam-driven scheme powered by a relativistic proton beam. Several such techniques were suggested earlier as a way of transferring the proton beam energy to other particles (electrons/positrons, muons, and pions) \cite{proton_klystron, coherent_acceleration} which cannot be accelerated in a circular machine either because of the prohibitively high energy loss due to the synchrotron radiation (electrons/positrons) or because of the short life-time of the unstable particles. The proton-driven plasma wakefield acceleration (PDPWFA) \cite{PDPWFA_nature} is a recently proposed method promising a GeV/m rate of acceleration to a TeV-scale energy in a single plasma stage. Plasma wakefields with GV/m amplitude correspond to the plasma wavelength around 1~mm, however the typical proton beam bunches in a storage ring have the longitudinal size in the order of 10~cm. In the original PDPWFA proposal \cite{PDPWFA_nature} the proton beam was assumed to be compressed by a factor of 1000 down to the longitudinal size of 0.1~mm. Such an extreme bunch compression is very challenging (although technically feasible), but the similarly high plasma wakefields can be excited resonantly with a sequence of sub-millimeter long microbunches produced from the long proton beam as a result of the beam self-modulation instability (SMI) \cite{EPAC98-806, SMI_PRL2010} in plasma. The self-modulated beam can drive the plasma wakefields for long distances \cite{density_step} even without the very strong external focusing needed in the case of the compressed proton bunch.

The AWAKE experiment \cite{AWAKE_PoP2014, Edda_EAAC2015, AWAKE_history} at CERN will be the first proof-of-principle demonstration of this technique using the self-modulation of the 400~GeV proton bunch from the SPS accelerator. A 10~m long plasma section based on the continuous flow of rubidium vapor at 200~$^{\circ}$C will be used. The selected plasma section design is already capable of creating plasma density profiles with a constant gradient along the whole 10~m long section \citep{Gennady_report, AWAKE_history}. This gradient naturally appears if the continuous flow of rubidium vapor through the orifices at the beam entrance and exit is unbalanced.

In general, plasma density gradients are well known to be useful for the control of electron injection and acceleration in the laser- and beam-driven plasma wakefield acceleration \cite{LWFA_gradient, Katsouleas_PRA_1986}. In this article we study the effect of a plasma density gradient on the acceleration of electrons in the plasma wake created by the self-modulating proton beam. The details of electron beam injection and acceleration in the uniform plasma were already given in \cite{on_axis_PoP2014}. More complicated longitudinal plasma density profiles were suggested earlier for the SMI-based PDPWFA \cite{Schroeder_PoP2012}. However we focus on the case of a constant plasma density gradient which is relevant for the AWAKE experiment. The process is studied numerically with the particle-in-cell version of 2d3v quasi-static code LCODE \cite{LCODE,PRST-AB6-061301,IPAC13-1238} assuming the cylindrical symmetry.

\section{The AWAKE experiment configuration}

\begin{figure}[t]
        \includegraphics[width = 0.48 \textwidth]{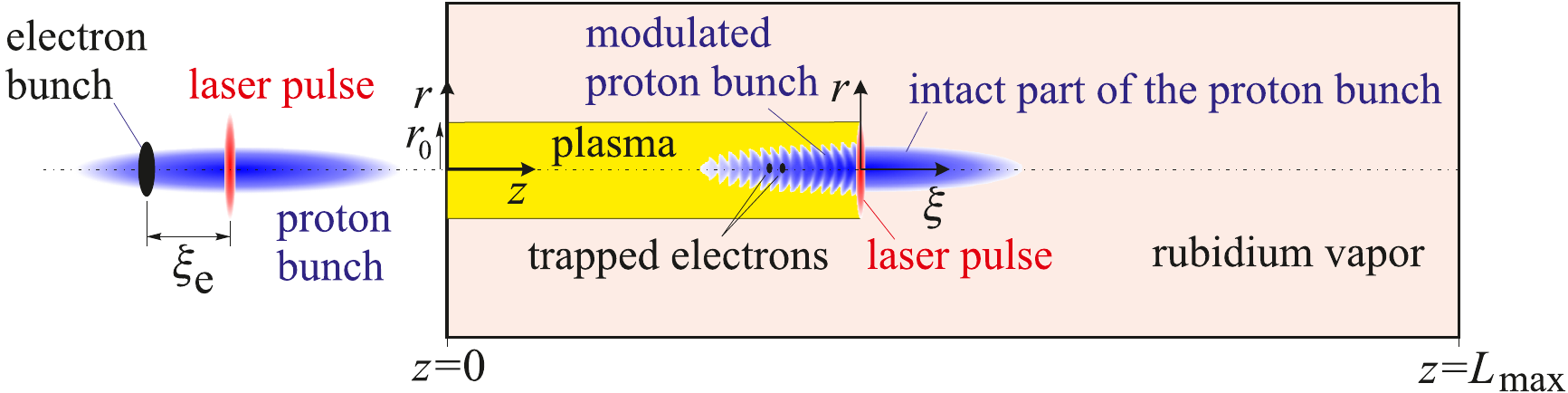}
        \caption{Geometry of the AWAKE experiment (not to scale). The beams are shown before (left) and inside (right) the plasma section. Baseline simulation parameters are the following: plasma density $n_0 = 7\cdot10^{14}~\mathrm{cm}^{-3}$ (plasma wavelength $\lambda_p = 1.26$~mm), plasma length $L_{\mathrm{max}} = 10$~m, plasma radius $r_0 = 1.4$~mm, number of protons $N_b=3\cdot10^{11}$, proton beam momentum $W_b = 400$~GeV/$c$, length $\sigma_{zb} = 12$~cm, radius $\sigma_{x,y~\mathrm{beam}} = 0.2$~mm and normalized emittance $\epsilon_{nb} = 3.6$~mm$\cdot$mrad; electron bunch energy $W_e = 16$~MeV, length $\sigma_{ze} = 1.2$~mm, radius $\sigma_{x,y} = 0.25$~mm, normalized emittance $\epsilon_{ne} = 2$~mm$\cdot$mrad, number of electrons $N_e = 1.25\cdot10^9$ and delay $\xi_e = 16.4$~cm.}
        \label{fig:layout}
\end{figure}

The AWAKE experiment will be conducted at CERN in the deep underground CNGS facility \cite{Edda_EAAC2015}. An LHC-type proton bunch of 400~GeV/$c$ momentum but higher intensity ($3\cdot10^{11}$ protons/bunch) is extracted from the SPS and sent towards a 10~m long rubidium vapor cell. A high power (2~TW) laser pulse, co-propagating and co-axial with the proton beam, is used to ionize the (initially neutral) rubidium gas in the plasma cell and also to generate a seed for the proton bunch self-modulation. A several millimeter long  bunch with $\sim10^{9}$ electrons at 10-20 MeV produced by the photo-injector serves as a witness beam and is accelerated in the wake of the proton bunches. Several diagnostics are installed downstream the plasma cell to measure the proton bunch self-modulation and the accelerated electron bunch properties. The geometry and the baseline parameters of the experiment are given in Fig.~\ref{fig:layout}.

\section{Effect of density gradient on the plasma wakefields}

The plasma wavelength is inversely proportional to the square root of the plasma density $n$:
\begin{equation}
\lambda_p = \sqrt{\frac{\pi}{n r_e}},
\end{equation}
where $r_e$ is the classical electron radius. Small change in the plasma density $\delta n$ results in the corresponding change in the plasma wavelength
\begin{equation}
\lambda_p + \delta \lambda_p = \sqrt{\frac{\pi}{(n + \delta n) r_e}} \approx \lambda_p \left( 1 - \frac{\delta n}{2n} \right).
\end{equation}
The phase shift of the wakefield corresponding to this plasma density change is
\begin{equation}
\label{eq:Delta_psi}
\Delta \psi = N\delta \psi = 2\pi N \frac {\delta \lambda_p} {\lambda_p} = -\pi N \frac{\delta n} {n},
\end{equation}
where $N$ is the number of plasma oscillations behind the ionizing laser pulse. Here for simplicity we consider free plasma oscillations driven by the seed pulse. Another cause of phase shifts is the evolution of the proton beam \cite{PoP22-103110}. During the SMI growth, the plasma wakefield phase and amplitude change even in the uniform plasma. In our case the typical phase shift due to the SMI is one plasma wavelength for 100 micro-bunches \cite{on_axis_PoP2014}. This shift accumulates over first 4 meters of the plasma. Only half of the wakefield period is decelerating for protons. Therefore, the number of proton beam micro-bunches driving the plasma wake resonantly in a non-uniform plasma can be estimated from $\Delta \psi \sim \pi$ and (\ref{eq:Delta_psi}) as
\begin{equation}
N \sim n/\delta n.
\end{equation}
Since the baseline AWAKE proton bunch has $\sigma_{zb}=12$~cm, the number of proton micro-bunches at the baseline density is around 100. We can expect that the typical plasma density gradients which can still support a significant wakefield amplitude are several percent over the entire length of the plasma section.

\begin{figure}[t]
        \includegraphics[width = 0.44 \textwidth]{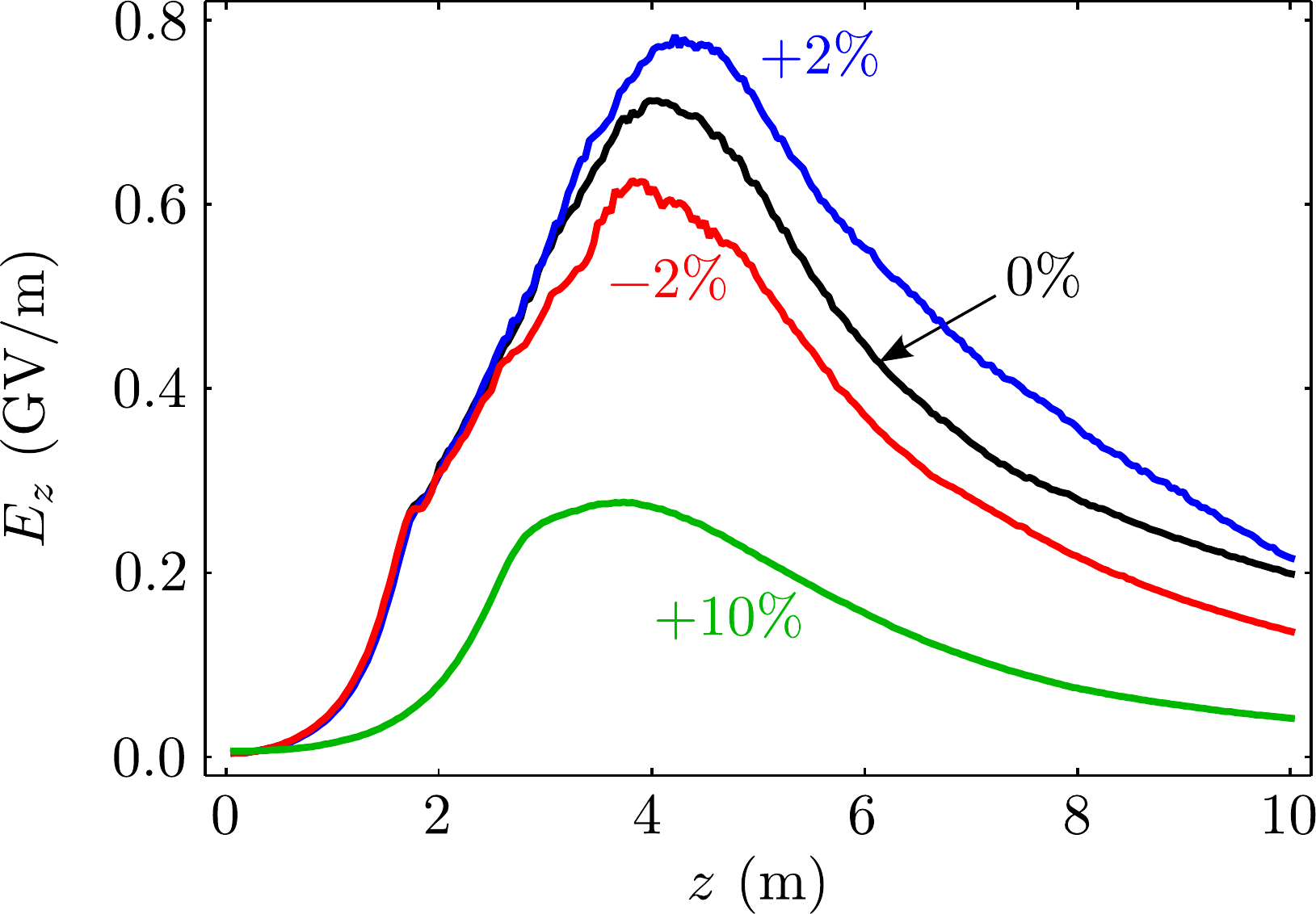}
        \caption{Maximum longitudinal electric field excited by the self-modulating proton beam in plasmas with different density gradients ($0\%$, $+2\%$, $-2\%$, and $+10\%$ over 10 m).}
        \label{fig:Ez_max}
\end{figure}

\begin{figure}[t]
        \includegraphics[width = 0.48\textwidth]{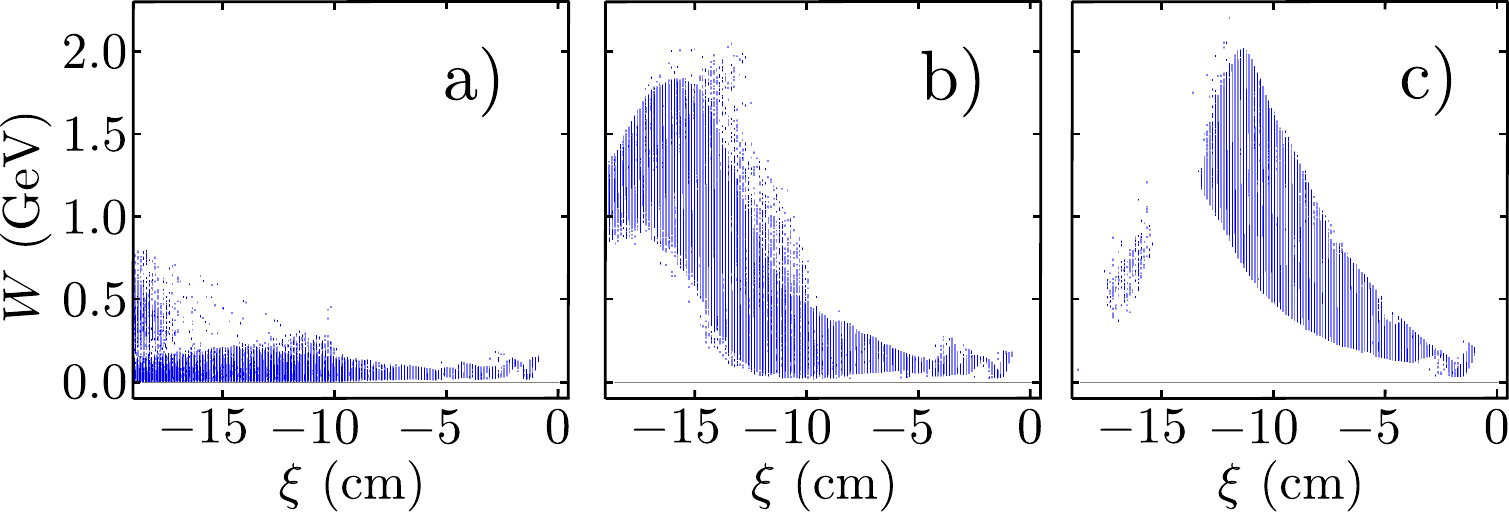}
        \caption{Effect of a small density gradient on acceleration of electrons in the wakefield of the self-modulating proton beam. In these simulations we use the long 16 MeV electron beam which initially overlaps with the proton beam over the whole simulation window. The electron beam charge in the simulations is negligibly small and it does not have any effect on the wakefield. The final electron energy vs distance to the laser pulse is shown in three cases: a) density gradient of $-2\%$ over 10~m b) uniform plasma c) density gradient of $+2\%$ over 10 m.}
        \label{fig:plus_minus_2_percent}
\end{figure}

The simulated amplitude of the longitudinal electric field excited by the self-modulating proton beam in plasmas with different density profiles is shown in Fig.~\ref{fig:Ez_max}. As expected, self-modulation of the proton beam with the baseline AWAKE parameters (Fig.~\ref{fig:layout}) produces wakefields with significant amplitudes for plasma density gradients up to several percent over 10~m (the 10\% gradient reduces the maximum field by a factor of three). However, the process of electron  acceleration is more sensitive to the density profile \cite{PoP20-013102}, especially to the sign of the density gradient which defines the phase velocity of the plasma wakefield. Fig.~\ref{fig:plus_minus_2_percent} shows the final accelerated electron energy as a function of the longitudinal position along the beam.

Density gradients offer a convenient way to control the phase of the plasma wakefield during the development of the self-modulation instability. In order to have a significant energy gain (\textgreater~1~GeV) in the uniform plasma, one has to inject electrons around 130 periods of plasma oscillations behind the laser pulse \cite{on_axis_PoP2014}, while in the case of a positive density gradient it is possible to shift the optimum injection delay closer to the laser pulse. This effect can be seen in Fig.~\ref{fig:plus_minus_2_percent}.

\begin{figure}[htb!]
        \includegraphics[width = 0.48 \textwidth]{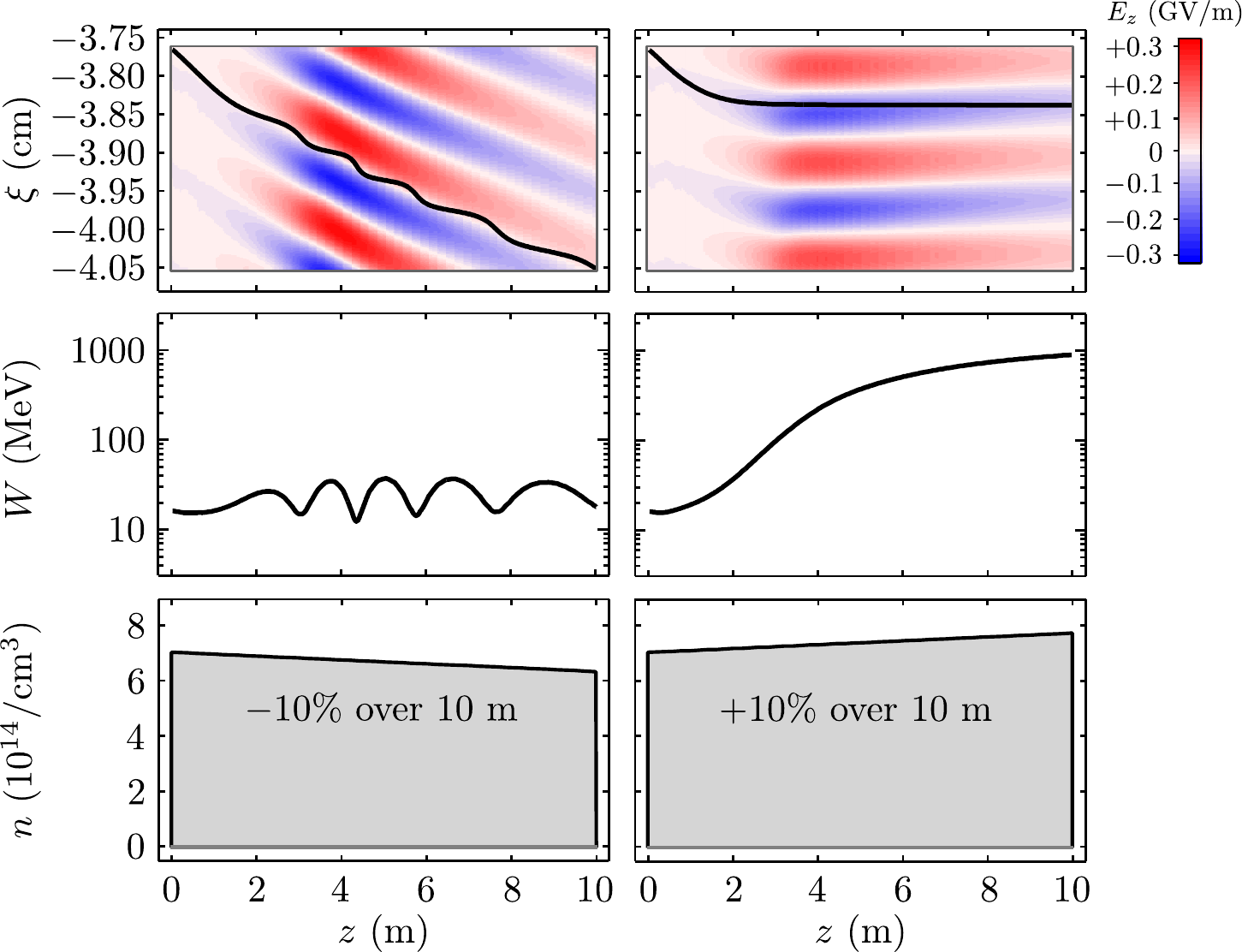}
        \caption{Longitudinal motion of an electron in the case of positive (right) and negative (left) plasma density gradient. The color map shows the amplitude of the longitudinal electric field on axis.}
        \label{fig:plus_minus_10_percent}
\end{figure}

The longitudinal motion of an individual electron as well as the wakefield phase and amplitude for positive and negative plasma density gradients are shown in Fig.~\ref{fig:plus_minus_10_percent}. As one can see, the amplitude of the electric field is approximately the same in both cases, however, the electron gains energy only in the case of the positive density gradient.

The negative density gradient results in a gradual increase of the plasma wavelength and corresponding continuous drift of the wakefield phase towards the tail of the bunch (i.~e., the phase velocity of the wakefield is slower than the speed of light). This limits the achievable energy, because, as soon as the injected electron gains some energy, it outruns the plasma wake and enters the decelerating (but still focusing) phase where it is decelerated. Such cycle of electron acceleration and deceleration can be repeated several times as one can see in the left side of Fig.~\ref{fig:plus_minus_10_percent}.

The positive density gradient makes the phase velocity of the plasma wake equal or slightly faster than the speed of light over a long part of the plasma section at some specific delay $\xi$ behind the laser pulse. Therefore it becomes possible for electrons to stay in phase with the wakefield and gain energy continuously until the end of the plasma section (right part of Fig.~\ref{fig:plus_minus_10_percent}). The steeper the plasma density gradient, the closer the optimal electron injection delay is to the laser pulse.

\begin{figure}[t]
        \includegraphics[width = 0.48\textwidth]{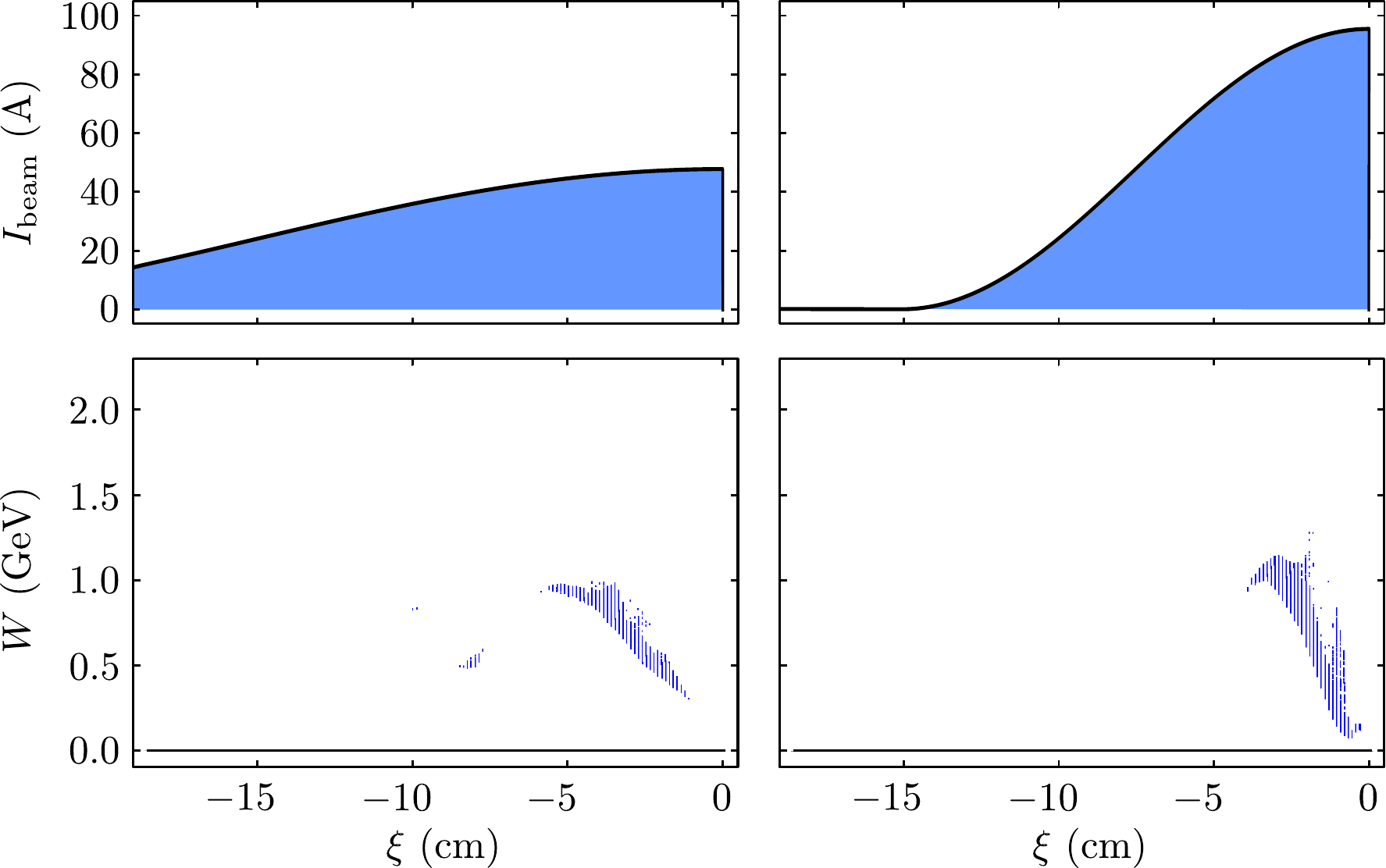}
        \caption{Longitudinal compression of the proton beam makes it possible to use steeper plasma density gradients. The simulation is similar to the one described in Fig.~\ref{fig:plus_minus_2_percent} (a long test electron beam initially co-propagates with the proton beam). The case of the baseline proton beam with $\sigma_{zb} = 12$~cm in the plasma with the density gradient of $+10\%$ over 10~m is shown on the left, while on the right the proton beam is compressed by a factor of two and the density gradient is two times steeper ($+20\%$ over 10~m). The top pictures show the proton beam current profile. The bottom pictures show the final electron beam energy vs distance to the laser pulse.}
        \label{fig:compressed}
\end{figure}

In general, the acceleration of electrons is easier to achieve by injecting particles closer to the laser pulse. The less micro-bunches are used for acceleration, the more stable and reproducible this process becomes with respect to various perturbations like small scale plasma non-uniformities or beam parameter variations from shot to shot. With a proton beam longitudinally compressed by a factor of two, the 20\% plasma density gradient (corresponding to the optimal injection at $\xi_e \approx -3$~cm) provides electron acceleration to more than 1~GeV (Fig.~\ref{fig:compressed}). In this case only 25 proton micro-bunches are required to accelerate electrons above 1~GeV. This is a factor of four reduction with respect to 100 micro-bunches necessary in the baseline case of uniform plasma and proton beam with $\sigma_{zb} = 12$~cm. Correspondingly, the requirement on small scale density uniformity can be four times relaxed.

\section{Conclusions}

Using numerical simulation we have shown that the plasma density gradients offer a simple and efficient way to control the development of the proton beam self-modulation instability and acceleration of electrons during this process.

For the baseline parameters of the AWAKE experiment it is possible to accelerate electrons to a high energy (\textgreater 1~GeV) in a plasma section with a density gradient of up to 10\% along 10~m. Using small positive gradients (below 10\% over 10~m) should improve the overall stability and reproducibility of electron acceleration experiment because it will rely on significantly lower number of micro-bunches to drive the plasma wakefield resonantly (optimum electron injection position shifts closer to the laser pulse). With longitudinally compressed proton bunches it is possible to operate at higher density gradients (20\% with 2x compression) –--- this can further improve the control over the acceleration process.

Even a small negative plasma density gradient ($-2$\% over 10~m for example) dramatically reduces the achievable electron energy because in this case the phase velocity of the plasma wake is always slower than the speed of light.

\section{Acknowledgements}

The authors thank AWAKE collaboration for fruitful discussions. The contribution of Novosibirsk team to this work is supported by The Russian Science Foundation, grant No.~14-12-00043. Some of the simulations are performed at the Siberian Supercomputer Center SB RAS.





\end{document}